\documentclass[11pt,preprint]{JHEP3}
 \preprint{ IFIC/05-30 $\qquad$ FTUV-05-1507 \\ July 15, 2005  \\ {\bf JHEP09 (2005) 064 } }

\title{Dirac equation for the supermembrane in a background with fluxes from a
component description of the D=11 supergravity--supermembrane interacting system}

\author {
{Igor A. Bandos$^{\dagger,\diamondsuit , \; \ast}$ and Jos\'e A. de
Azc\'arraga$^{\dagger , \;}$\footnote{ e-mails: bandos@ific.uv.es,
j.a.de.azcarraga@ific.uv.es}}

\vspace{1cm}

{\it $^{\dagger}$ Departamento de F\'{\i}sica Te\'orica, Univ.~de Valencia and IFIC
(CSIC-UVEG), 46100-Burjassot (Valencia), Spain

$^{\diamondsuit}$Institute for Theoretical Physics, NSC ``Kharkov Institute of Physics
and Technology'',  UA61108, Kharkov, Ukraine }

}

\abstract{We present a simple derivation of the `Dirac' equation for the supermembrane
fermionic field in a D=11 supergravity background with fluxes by using a complete but
gauge--fixed description of the supergravity--supermembrane interacting system
previously developed. We also discuss the contributions linear in the supermembrane
fermions --the Goldstone fields for the local supersymmetry spontaneously broken by the
superbrane-- to the field equations of the supergravity--supermembrane interacting
system. The approach could also  be applied to more complicated dynamical systems such
as those involving the M5--brane and the D=10 Dirichlet branes.
 }

\begin{document}

\maketitle

\section{Introduction}

There has been a growing interest in the fermionic equations for
superbranes in a supergravity background with fluxes (see
\cite{Dima+RK05,VP+2005} and
\cite{BBS95,deW+PP98,HarveyMoore99,BranesSG} for earlier papers)
as these are needed to study nonperturbative effects in string
theory. To find such equations, one takes the super--$p$--brane
action in curved superspace and expands it in powers of the
fermionic coordinate function $\hat{\theta}^{\check{\beta}}(\xi)$
\cite{BBS95,deW+PP98,HarveyMoore99,BranesSG,VP+2005} or proceeds
in the same manner directly from the `superfield' fermionic
equation itself \cite{Dima+RK05}. To make such a decomposition one
uses the Wess--Zumino (WZ) gauge for a superfield supergravity
background plus the superspace supergravity constraints. For
$D=10, 11$ these superspace constraints  imply the `free'
supergravity equations of motion, without any contribution from
the superbrane source. Hence the `Dirac' equation for
$\hat{\theta}(\xi)$, as derived in \cite{Dima+RK05} from
superembedding approach \cite{6*} and in
\cite{deW+PP98,HarveyMoore99,BranesSG,VP+2005} from the ${\cal
O}(\hat{\theta}(\xi)^{\wedge 2})$ actions for M2 and various
super--Dp-branes, apparently involves a gravitational and gauge
field background that satisfies the `free' supergravity equations
without any contribution from the superbrane. Thus the consistency
of the results obtained in this approach, although widely
believed, is not manifest. A way to check this consistency would
be to re-derive the same equations from a system of equations
providing a fully dynamical description of the
supergravity--super-$p$-brane interacting system by using a well
defined approximation.

As, by definition, a super-$p$-brane is a $p$-brane moving in superspace, a complete
system of equations, including those for the supergravity fields with contributions
from the super-$p$-brane, could be derived from  the sum of the superbrane action and a
{\it superfield} action for supergravity. Such an action can be studied in lower
dimensions (see \cite{BAIL03,BAIL-BI,BI03}), but a {superfield} action for $D=10,11$
supergravity is unknown. This made difficult the study  the $D=11$
supergravity--M-brane and $D=10$ type II supergravity--D$p$-brane interacting systems,
the most interesting ones in an M--theoretical perspective.

This difficulty may be  overcome  by using the gauge--fixed description
\cite{BAIL03,BAIL-BI,BI03} provided by the sum of the spacetime (component) action for
supergravity (without auxiliary fields) and that for a {\it bosonic} brane, as given by
the purely bosonic limit of the superbrane action. From the point of view of the
superfield formulation of the interacting system (hypothetical for $D=10, 11$) the
gauge is provided  by the conditions of the WZ gauge for the supergravity superfields
plus the condition $\hat{\theta}^\alpha(\xi)=0$ for the superbrane coordinate
functions. The resulting gauge--fixed description  is complete \cite{BAIL-BI} in the
sense that it contains gauge--fixed versions of all the dynamical equations of motion
of the interacting system, including a `fermionic equation for the bosonic brane'
\cite{BdAI01}. This equation, which formally coincides with the leading component of
the superfield fermionic equation for the superbrane in a superspace supergravity
background, appears in this component scheme as a selfconsistency condition for the
bulk gravitino equations. Note that the `fermionic equation for bosonic brane' is
actually  a non-dynamical `boundary' condition for the gravitino on the brane
worldvolume $W^{p+1}$. However, we will see how this algebraic equation allows us to
obtain the superbrane fermionic field dynamics (the `Dirac equation'), which is hidden
in it.

The above gauge--fixed action for the supergravity--superbrane interacting system can
be derived from the superfield description in the dimensions where a superfield
supergravity action exists (see \cite{BAIL03,BI03} for $D=4$, $N=1$ interacting
systems). In the general case its form may be also deduced if one assumes the existence
of a superfield supergravity action and exploits its defining properties
\cite{BAIL-BI}. Then one concludes that, whether a superfield description of a
supergravity--superbrane interacting system exists or not, the description of this
system by means of the sum of the spacetime component supergravity action without
auxiliary fields and the action of the `limiting' bosonic brane (obtained by taking the
purely bosonic, $\hat{\theta}^\alpha(\xi)=0$, `limit' of the supermembrane) does exist.
Such an action preserves one--half of the local supersymmetry \cite{BdAI01}
characteristic of the pure supergravity action. This one-half of the local
supersymmetry reflects the $\kappa$--symmetry of the original superbrane action while
the existence of a non--preserved one-half reflects the spontaneous breaking of the
local supersymmetry by the superbrane.

In this paper we show that this complete but gauge--fixed description of the
supergravity--superbrane interacting system may be used, despite it corresponds to the
$\hat{\theta}^\alpha(\xi)=0$ gauge for the  superbrane variables, to reproduce the
superbrane fermionic equation {\it i.e.}, the dynamical Dirac equation for the
fermionic field $\hat{\theta}^\alpha(\xi)$, in  a supergravity background with fluxes.
This is related to the known fact that the superbrane fermionic coordinate functions
are the Goldstone fields for the supersymmetry spontaneously broken by the superbrane.
We also discuss the possibility of using  the Goldstone nature of the
$\hat{\theta}^\alpha(\xi)$'s  to find their  lower order (${\cal
O}(\hat{\theta}^{\wedge k})$) contributions to the bulk supergravity equations, {\it
i.e.} to search for a lower--order approximation in $\hat{\theta}^\alpha(\xi)$ to the
system of interacting equations that would possess full local supersymmetry (not just
the  one half preserved by the gauge--fixed description of \cite{BAIL-BI,BAIL03,BI03})
in the same (actually, ${\cal O}(\hat{\theta}^{\wedge (k \! -\! 1)})$) approximation.

For definiteness we consider here the case of the $D=11$ supergravity--supermembrane
({\it SG--M2})  interaction, although the method could be applied to other systems like
the {\it SG--M5} one involving the M5--brane, or  the {\it SG-Dp} system, with $D=10$
type II Dirichlet $p$--branes.

\setcounter{equation}0

\section{Supergravity interacting with a bosonic membrane as a gauge--fixed description
of the D=11 supergravity--supermembrane ({\it SG--M2}) interacting system and its
properties}

\subsection{ D=11 Supermembrane in the on-shell superfield
supergravity background}

The supermembrane action in a supergravity background is \cite{BST87}
\begin{eqnarray} \label{SM2ssp}
{S}_{_{M2}}[\hat{E}^a, \hat{A}_3]=
 \int\limits_{W^{3}}
\left(d^3\xi {1\over 2} \sqrt{|g(\xi)|} - \hat{A}_3 \right)=
 \int\limits_{W^{3}}
\left({1\over 3!} \hat{E}^a \wedge \hat{\star} \hat{E}^a - {A}_3(\hat{Z})\right) ,
\qquad
\end{eqnarray}
where the pull--backs to the supermembrane worldvolume $W^3$
\begin{eqnarray}
\nonumber
\label{hEa} \hat{E}^a:= d\hat{Z}^M(\xi) E_M{}^a
(\hat{Z}(\xi)) =: d\xi^m \hat{E}_m^a
\end{eqnarray}
and
\begin{equation}
\quad \hat{A}_3 = {1\over 3!} d\hat{Z}^{M_3} \wedge d\hat{Z}^{M_2}
\wedge d\hat{Z}^{M_1} A_{M_1 M_2 M_3} (\hat{Z}(\xi))
\end{equation}
of the bosonic supervielbein $E^a= dZ^M E_M{}^a(Z)$ and the 3--superform $A_3=
{{1}\over{3!}} d{Z}^{M_3} \wedge d{Z}^{M_2} \wedge d{Z}^{M_1} A_{M_1 M_2 M_3} ({Z})$ of
the superspace formulation \cite{CF80,BH80} of $D=11$ supergravity \cite{CJS} are
denoted by a caret. These are obtained by replacing the superspace coordinate $Z^M=
(x^\mu, \theta^{\check{\alpha}})$ by the coordinate functions $\hat{Z}^M(\xi)=
(\hat{x}^\mu(\xi), \hat{\theta}^{\check{\alpha}}(\xi))$ that `locate' the worldvolume
$W^3$ as a hypersurface in superspace, $Z^M=\hat{Z}^M(\xi)$. The worldvolume Hodge star
operator $\hat{\star}$ in (\ref{SM2ssp}) is defined  by
\begin{eqnarray} \label{hHodge}
\hat{\star} \, \hat{E}^a = {1\over 2} d\xi^{n} \wedge d\xi^m
\epsilon_{mnk}\sqrt{|g(\xi)|}\; g^{kl}
\partial_l \hat{Z}^M\, E_M{}^{a}(\hat{Z})\; ,
\end{eqnarray}
where $g_{mn}(\xi)$ is the induced metric on $W^3$,
\begin{eqnarray}\label{ind-g}
g_{mn}(\xi):=  \hat{E}_m^{a} \, \hat{E}_{n\, a} :=
\partial_m \hat{Z}^M \partial_n \hat{Z}^N\, E_N{}^{a}(\hat{Z}) \,
E_{M\, a}(\hat{Z})\; , \quad |g(\xi)|:= |det(g_{mn}(\xi))| \; . \quad
\end{eqnarray}

 The supermembrane  equations of motion
\begin{eqnarray}\label{SEqmX}
&& \hat{D}(\hat{\star} \hat{E}_a)= -  {1\over 3} \hat{E}^{d}\wedge\hat{E}^{c} \wedge
\hat{E}^{b} \, F_{abcd}(\hat{Z}) - {1\over 2} \hat{E}^{d}\wedge \hat{E}^{\alpha} \wedge
\hat{E}^{\beta} \Gamma_{ab \, \alpha\beta} \,   \; , \qquad
 \\
\label{SEqmTh} &&  \hat{\star} \hat{E}^a \wedge \hat{E}^\beta
\left(\Gamma_a(1-\bar{\bar{\gamma}})\right)_{\beta \alpha} = 0 \; \qquad
\Leftrightarrow \qquad \hat{E}_m^\beta
\left(\hat{\Gamma}^m(1-\bar{\bar{\gamma}})\right)_{\beta \alpha} = 0 \; , \quad
\end{eqnarray}
where $D$ is the standard covariant derivative involving the spin connection,
$\hat{E}_m^{\, A}:= \partial_m\hat{Z}^M E_{M}^{\, A}(\hat{Z})$, $\hat{\Gamma}^m:=
g^{mn}(\xi) \hat{E}_n^{\; a}\, \Gamma_a$ and
\begin{eqnarray}\label{bgam-SSP}
\bar{\bar{\gamma}} = & {i\over 3!\sqrt{|g(\xi) |}}\; \epsilon^{mnk}
\partial_k \hat{Z}^K
\partial_n \hat{Z}^N
\partial_m \hat{Z}^M  E_M^{\;\; a}(\hat{Z}) E_N^{\;\; b}(\hat{Z}) E_K^{\;\; c}(\hat{Z})
 & \!\Gamma_{abc}\; , \qquad
\end{eqnarray}
are obtained \cite{BST87} making use of the superspace supergravity constraints
\cite{CF80,BH80}
\begin{eqnarray}
\label{cTa=} T^a&:=& DE^a := dE^a - E^b\wedge \omega_b{}^a  = -i E^\alpha\wedge E^\beta
\Gamma^a_{\alpha\beta} \qquad ,
\\ \label{cdA3=}
dA_3 &=& {1\over 4} E^\alpha\wedge E^\beta \wedge E^b \wedge E^a \Gamma_{ab \,
\alpha\beta} + F_4(Z)\; , \qquad F_4(Z):= {1\over 4!} E^{a_4} \wedge \ldots \wedge
E^{a_1} F_{a_1 \ldots a_4}(Z) \; . \qquad
\end{eqnarray}
These are known to be {\it on--shell} constraints {\it i.e.}, they include the
equations of motion for the physical spacetime or `component' fields $e_\mu^a(x)$,
$\psi_\mu^\alpha(x)$, $A_{\mu\nu\rho}(x)$,
\begin{eqnarray}\label{e=E}
e_\mu^a(x)\; &:& \qquad e^a(x):=dx^\mu e_\mu{}^a(x) = E^a(Z)\vert_{\theta = 0,\,
d\theta = 0 } \quad \\ \label{f=E} \psi_\mu^\alpha(x)\; &:& \qquad
 \psi^\alpha(x):= dx^\mu  \psi_\mu^{\, \alpha}(x)= E^\alpha(Z)\vert_{\theta
 = 0\, d\theta = 0}
 \\ \label{A=A} A_{\mu\nu\rho}(x)&:&  \qquad
 A_3(x):= {1/ 3!}\; dx^\mu \wedge dx^\nu \wedge dx^\rho A_{\mu\nu\rho}(x)
 = A_3(Z)\vert_{\theta = 0, \,  d\theta = 0}\; ,
\end{eqnarray}
among their consequences. For our present discussion it is important to note that these
are  `free' supergravity equations in the sense that they do not contain  any source
contribution from the supermembrane.

The on--shell supergravity constraints (\ref{cTa=}), (\ref{cdA3=}) are also necessary
conditions for the $\kappa$--symmetry of the supermembrane in a curved superspace
background \cite{BST87}. This local fermionic symmetry, first found  in a superparticle
context in \cite{dA+L-S}, manifests itself by the presence of the projector
$(1-\bar{\bar{\gamma}})$ in the fermionic equations (\ref{SEqmTh}). Its explicit form
is given by
\begin{equation}\label{kappa}
\delta_\kappa \hat{Z}^M(\xi) = (1+ \bar{\bar{\gamma}})^\alpha{}_\beta \, \kappa^\beta
(\xi) \, E_{\alpha}{}^M(\hat{Z}) \; .
\end{equation}
When the supergravity background superfields obey the Wess--Zumino (WZ) gauge
conditions
\begin{eqnarray}\label{WZg}
i_\theta E^a &:=& \theta^{\check{\beta}} E_{\check{\beta}}{}^a (x ,\theta )  =0 \; ,
\qquad  \nonumber \\  i_\theta E^\alpha   &:=&  \theta^{\check{\beta}}
E_{\check{\beta}}{}^\alpha (x ,\theta ) = \theta^{\check{\beta}} \delta
_{\check{\beta}}{}^\alpha =: \theta^\alpha \; , \\ &&  \nonumber i_\theta A_3(Z)=0 \; ,
\qquad i_\theta \omega^{ab}(Z)=0 \;  \qquad
\end{eqnarray}
(see \cite{BAIL03} and refs. therein; see also \cite{Tsimpis04})\footnote{$i_\theta$ is
a shorthand for the contraction with the vector field $\theta^{\check{\alpha}}{\partial
\over \partial \theta^{\check{\alpha}}}$.} the index of the Grassmann coordinate
$\theta$ and of the coordinate function $\hat{\theta}(\xi)$ is identified with the
spinor index and, furthermore, one can extract from (\ref{kappa}) the
$\kappa$--symmetry transformations for the $\hat{\theta}^\alpha$ variable,
\begin{equation}\label{kappaTH}
\delta_\kappa \hat{\theta}^\alpha(\xi) = (1+ \bar{\bar{\gamma}})^\alpha{}_\beta \,
\kappa^\beta (\xi) \, + {\cal O} (\hat{\theta}) = (1+ {\bar{\gamma}})^\alpha{}_\beta \,
\kappa^\beta (\xi) \, + {\cal O} (\hat{\theta}) \; ,
\end{equation}
where in Eq. (\ref{kappaTH}) $\bar{\gamma}= \bar{\bar{\gamma}}\vert_{\hat{\theta}=0}$.

The name {\it Dirac equation for the superbrane} is usually given
\cite{Dima+RK05,VP+2005} for (an approximation to) the equation of motion for the
superbrane fermion $\hat{\theta}(\xi)$ in a {\it spacetime} supergravity background. We
will also call it below {\it superbrane fermionic equation}. In the standard approach
to derive this equation \cite{deW+PP98,HarveyMoore99,BranesSG,Dima+RK05,VP+2005} one
considers the action (\ref{SM2ssp}) \cite{deW+PP98,HarveyMoore99,BranesSG,VP+2005} or
\cite{Dima+RK05} the superfield fermionic equation (\ref{SEqmTh}) for the on--shell
supergravity background taken in the WZ gauge and expands it in powers of
$\hat{\theta}(\xi)$ keeping the lower orders in $\hat{\theta}(\xi)$; the first order is
usually considered to be sufficient. Then one uses the $\kappa$--symmetry (\ref{kappa})
to gauge away half ($16$ out of $32$) of the $\hat{\theta}(\xi)$ components  to retain
only the physical supermembrane fermions.

The fact that both the very derivation of the superfield fermionic equation
(\ref{SEqmTh}) and its decomposition in powers of $\hat{\theta}(\xi)$ makes an
essential use of  the {\it on--shell} superspace supergravity constraints, which cannot
incorporate any supermembrane source contribution, makes the consistency of the
standard background superfield approach
\cite{deW+PP98,HarveyMoore99,BranesSG,Dima+RK05,VP+2005} not obvious.  The check of its
consistency is one of the motivations of this paper.

\subsection{On the properties of a (hypothetical) superfield Lagrangian
description of the D=11 supergravity--superbrane interaction}

A complete, supersymmetric description of the {\it SG--M2} interaction would be
provided by the sum
\begin{eqnarray}\label{SSGZ+SM2}
\mathbf{S}_{SG-M2}= \mathbf{S}_{SG} [E^a, E^\alpha , A_3(Z)] + {S}_{M2}[\hat{E}^a ,
{A}_3(\hat{Z})]\;
\end{eqnarray}
of the supermembrane action (\ref{SM2ssp}) and the hypothetical superfield action for
$D=11$ supergravity $\mathbf{S}_{SG}[E^a, E^\alpha , A_3(Z)]$. This action is not known
and it is not even clear whether it exists. Nevertheless, if exists, such a
supergravity action would possess certain properties. In particular, it would be
invariant under arbitrary changes of the superspace coordinates, {\it i.e.}
superdiffeomorphisms $\delta_{sdiff}$. The same is true of the full interacting action
(\ref{SSGZ+SM2}) provided \cite{BAIL03,BAIL-BI} that the transformations of the {\it
coordinate functions} of superbrane, $\hat{Z}^M(\xi)= (\hat{x}^\mu (\xi),
\hat{\theta}^{\check{\alpha}}(\xi))$ are given by the pull--backs $\hat{b}^M=
b^M(\hat{Z}(\xi))$ to $W^3$  of the superspace diffeomorphism parameters ${b}^M({Z})$,
{\it i.e.}
\begin{eqnarray}\label{sdiffZ}
\delta_{sdiff} {Z}^M = b^M({Z}) \; ,  \quad \delta_{sdiff} \hat{Z}^M(\xi) =
b^M(\hat{Z}(\xi)) \; .
\end{eqnarray}
Eq. (\ref{sdiffZ}) implies, in particular,
\begin{eqnarray}\label{sdiffTh}
\delta_{sdiff}{\theta}^{\check{\alpha}}= b^{\check{\alpha}}(Z)\; , \qquad
\delta_{sdiff} \hat{\theta}^{\check{\alpha}}(\xi) = {b}^{\check{\alpha}}(\hat{Z}(\xi))
\; .
\end{eqnarray}
Clearly, the transformations $\delta_{sdiff}{Z}^{M}= b^{M}(Z)$ cannot be used to set
the fermionic {\it coordinates} $\theta^\alpha$ equal to zero since such a
transformation would have a vanishing superdeterminant and, hence, would not be a
superdiffeomorphism. However, in contrast, the transformations (\ref{sdiffTh}) {\it
can} be used to make the fermionic coordinate {\it  functions} $\hat{\theta}^\alpha
(\xi)$ vanishing, {\it i.e.} one can fix  the gauge \cite{BAIL03,BAIL-BI}
\begin{eqnarray}
\label{thGAUGE0}  \hat{\theta}^\alpha (\xi)=0 \; ,
\end{eqnarray}
which might be considered the analogue to the `unitary gauge' of the Higgs model.

Another expected property of the hypothetical superfield interacting action
(\ref{SSGZ+SM2}) is that, in addition to the superspace diffeomorphism gauge symmetry
(Eqs. (\ref{sdiffZ}), (\ref{sdiffTh})),  it would possess a local $16$--parametric
fermionic $\kappa$--symmetry $\delta_\kappa$ acting on the supermembrane variables
$\hat{Z}^M(\xi)$ only. It is also plausible to assume that such a $\kappa$--symmetry
would be characterized by Eq. (\ref{kappa}) with some superfield projector
$1/2(1+\bar{\bar{\gamma}})$, $\bar{\bar{\gamma}}\equiv  \bar{\bar{\gamma}}(Z)$. Thus
the set of fermionic gauge symmetries of the action would contain
$\delta_{gauge}=\delta_{sdiff} +\delta_{\kappa}$. These transformations act on
$\hat{\theta}^{\alpha}(\xi) $ as
\begin{eqnarray}\label{sdif+kap}
\delta_{gauge} \hat{\theta}^{\alpha}(\xi) = b^{\alpha}(\hat{Z}(\xi)) + \delta_{\kappa}
\hat{\theta}^{\alpha}(\xi) =  - \varepsilon^{\alpha}(\hat{x})  + {\cal O}
(\hat{\theta}(\xi)) + \delta_{\kappa}
 \hat{\theta}^{\alpha}(\xi) \; ,
\end{eqnarray}
where the leading component of the superfield superdiffeomorphism parameter has been
denoted by $- \varepsilon^{\alpha}(\hat{x})$,
\begin{eqnarray}
b^{\alpha}(\hat{Z}(\xi))
 = - \varepsilon^{\alpha}(\hat{x}) + {\cal O} (\hat{\theta}(\xi)) \;
\end{eqnarray}
to identify $\varepsilon^{\alpha}(\hat{x})$ with the spacetime local supersymmetry
parameter.

Irrespective of the details of the superspace formulation of supergravity, the WZ gauge
(\ref{WZg}) can be fixed on the supergravity superfields (see {\it e.g.} \cite{CF80},
\cite{BAIL03} and refs. therein) by using superdiffeomorphism symmetry (\ref{sdiffZ})
and the superspace structure group symmetry, $SO(1,10)$ in the present case. The WZ
gauge is then preserved by a certain combination of the superdiffeomorphism and the
superspace local Lorentz group transformations expressed in terms of a number of
independent parameters, $\epsilon^{\alpha}({x})$ of the spacetime local supersymmetry,
$b^\mu(x)$ of spacetime diffeomorphisms and  $L^{ab}(x)$ of spacetime local Lorentz
transformations. In the WZ gauge the transformations of the fermionic coordinate
function of the superbrane, $\hat{\theta}^{\alpha}(\xi)$, read
\begin{eqnarray} \label{sdif-kap} \delta_{\kappa}
 \hat{\theta}^{\alpha}(\xi) = (1 + \bar{\bar{\gamma}}(\hat{Z}))^{\alpha}{}_\beta
 \kappa^\beta(\xi) = (1 + {\bar{\gamma}}(\hat{x}))^{\alpha}{}_\beta
 \kappa^\beta(\xi) + {\cal O} (\hat{\theta}(\xi)) \; ,
\end{eqnarray}
where $\bar{\gamma}\equiv \bar{\gamma}(x)$ is the leading component of
$\bar{\bar{\gamma}}\equiv \bar{\bar{\gamma}}(\hat{Z})$
 in the
$\kappa$--symmetry projector,  $\bar{\bar{\gamma}}(\hat{Z})\vert_{\hat{\theta}=0}=
\bar{\gamma}(x)$.  Substituting (\ref{sdif-kap}) for $\delta_{\kappa}
 \hat{\theta}^{\alpha}(\xi)$ in (\ref{sdif+kap}) one finds
\begin{eqnarray}\label{sdif+kap2}
\delta_{gauge} \hat{\theta}^{\alpha}(\xi) =
 - \varepsilon^{\alpha}(\hat{x}) + (1 +
{\bar{\gamma}})^{\alpha}{}_\beta
 \kappa^\beta(\xi) + {\cal O}
(\hat{\theta}(\xi))\; ,
\end{eqnarray}
Eq. (\ref{sdif+kap2}) exhibits, first of all,  the Goldstone nature of the superbrane
fermionic coordinate functions $\hat{\theta}^{\alpha}(\xi)$: $\;$
$\hat{\theta}^{\alpha}(\xi)$ are the Goldstone fermions corresponding to the
supersymmetry spontaneously broken by the superbrane (see \cite{H+L+Polchinski86},
\cite{BAIL03,BAIL-BI} and refs. therein). In the supergravity--superbrane interacting
system this supersymmetry is the {\it spacetime} local gauge symmetry which can be used
to remove the Goldstone field by fixing the gauge (\ref{thGAUGE0}). Secondly, Eq.
(\ref{sdif+kap2}) makes transparent that the spontaneous breaking of the local
supersymmetry by superbrane is partial. Indeed, the simple observation
\begin{eqnarray}\label{susyKap2}
\hat{\theta}^{\alpha}(\xi) =0  & \Rightarrow & 0= \delta_{gauge}
\hat{\theta}^{\alpha}(\xi) \vert_{\hat{\theta}(\xi) =0}= -
\varepsilon^{\alpha}(\hat{x}) + (1 +
{\bar{\gamma}})^{\alpha}{}_{\beta}\kappa^\beta(\xi)
  \;  \quad \end{eqnarray}
implies that the gauge (\ref{thGAUGE0}) is preserved by a local supersymmetry of
parameter $\varepsilon(x)$ whose pull--back to the brane is restricted by being of the
form
\begin{eqnarray} \label{susy1/2}
 && \hat{\varepsilon}^{\alpha}(\xi) := \varepsilon^{\alpha}(\hat{x}) =
(1 + {\bar{\gamma}}(\hat{x}))_{\beta}\kappa^\beta(\xi) \; .
\end{eqnarray}

\subsection{Gauge--fixed description of the {\it SG--M2} interacting system and its properties}

Hence, as shown in \cite{BAIL03,BAIL-BI} and discussed above, in a hypothetical
superfield description (\ref{SSGZ+SM2}) of the supergravity--superbrane interacting
system the gauge
\begin{eqnarray}
\label{thGAUGE}  \hat{\theta}^\alpha (\xi)=0 \;
\end{eqnarray}
(Eq. (\ref{thGAUGE0})) and the WZ gauge (\ref{WZg}) may be fixed simultaneously. In the
WZ gauge the integration over the Grassmann superspace coordinates $\theta^\alpha$ in
such a superfield action $\mathbf{S}_{SG}[E^a, E^\alpha , A_3(Z)]$ would produce a
component spacetime supergravity action involving a (hypothetical) set of auxiliary
fields.  By definition, these auxiliary fields would satisfy algebraic equations which,
used in the supergravity action, would lead to the standard supergravity action (in our
case that of \cite{CJS}) involving only  the physical fields of the supergravity
multiplet. This action is invariant under  the local supersymmetry the algebra of which
closes on--shell.

Notice that the auxiliary fields would be contained in the higher order components of
$E_M^{\; A}(Z)$, $A_{MNK}(Z)$ (and, perhaps, in some additional auxiliary superfields).
The leading ($\theta=0$) components of the $E_M^{\; A}(Z)$ and $A_{MNK}(Z)$ superfields
in the WZ gauge are either zero, unity, or, in the case of $E_\mu^A(Z)$ and
$A_{\mu\nu\rho}(Z)$, determine the physical fields $e_\mu^a(x)$, $\psi^\alpha(x)$ and
$A_{\mu\nu\rho}(Z)$ of the Cremmer--Julia--Scherk (CJS) supergravity multiplet
\cite{CJS} (Eqs. (\ref{e=E}), (\ref{f=E}) and (\ref{A=A})). As a result, in the gauge
defined by $\hat{\theta}(\xi)=0$ (Eq. (\ref{thGAUGE})) plus the WZ gauge (Eqs.
(\ref{WZg})), the supermembrane action (\ref{SM2ssp}) reduces to the action $S^0_{M2}$
of a purely {\it bosonic membrane} coupled to the {\it physical bosonic fields} of the
supergravity multiplet only; neither the gravitino nor the auxiliary fields enter the
membrane part of the gauge--fixed interacting action.

Hence, in the supergravity part of such a gauge--fixed action for the {\it SG--M2}
interacting system one may remove the auxiliary fields through their  algebraic
equations in the same manner that one would do for the (also hypothetical) pure
supergravity action with auxiliary fields. As a result one would arrive at the
following gauge--fixed action for the {\it SG--M2} interacting system
 \cite{BAIL-BI,BAIL03} (see also \cite{BdAI01})
\begin{eqnarray}
\label{SSG+M2} S^0_{SG-M2}= S_{SG}[e^a , \psi^\alpha , A_3] + S^0_{M2}= \int_{M^{11}}
{\cal L}_{11}[e^a , \psi^\alpha , A_3]  + \int_{W^{3}} \left({1\over 3!} \hat{e}^a
\wedge \hat{\ast} \hat{e}_a - {A}_3(\hat{x})\right) \; ,
\end{eqnarray}
where $S_{SG}= S_{SG}[e^a , \psi^\alpha , A_3]$ is the standard CJS action for D=11
supergravity \cite{CJS} and the second term is the action for a purely bosonic brane
where the relative coefficient between its two terms is fixed (for a given supergravity
action $S_{SG}[e^a , \psi^\alpha , A_3]$ invariant under definite supersymmetry
transformations) since $S^0_{M2}=S_{M2}[{e}^a(\hat{x}), {A}_3(\hat{x})]$ is the bosonic
limit of the M2--superbrane action $S_{M2}[{E}^a(\hat{Z}), {A}_3(\hat{Z})]$
\cite{BST87} of Eq. (\ref{SM2ssp}).

The following properties \cite{BdAI01,BAIL-BI} of the gauge--fixed action
(\ref{SSG+M2}) will be important
\begin{itemize}
\item{{\bf 1)}} The gauge--fixed description (\ref{SSG+M2}) of the
supergravity--superbrane interacting system (\ref{SSGZ+SM2}) is {complete} in the sense
that it produces a gauge--fixed version of {all} the dynamical equations that would be
obtained from a possible superfield action,  including the {\it `fermionic equation for
bosonic brane'} \cite{BdAI01,BAIL03}, which is given by an algebraic condition on the
pull--back $\hat{\psi}^\alpha:= d\xi^m \hat{\psi}_m^\alpha(\xi) $ of the gravitino to
$W^3$. It states that a projection of a gamma--trace of $\hat{\psi}_m^\alpha$ vanishes,
{\it i.e.} that
\begin{eqnarray}
\label{XIM2-c} \hat{\psi}_m \hat{\Gamma}^m(1-\bar{\gamma})  = 0 \; \qquad
\end{eqnarray}
(see Sec. 2.3.2), where
\begin{eqnarray}
\hat{\psi}_m^\beta := \hat{e}_m^{\; a}(\xi) {\psi}_a^\beta(\hat{x}(\xi)) =
\partial_m \hat{x}^\mu \psi_{\mu}^{\alpha}(\hat{x})
\, , \qquad  \hat{e}_m^{\; a} :=
\partial_m\hat{x}^\mu(\xi) {e}_\mu{}^{a}(\hat{x}(\xi)) \, , \qquad \nonumber \\
\hat{\Gamma}^n_{\alpha\beta}:= g^{nm}(\xi)\hat{e}_m^{\; a}(\xi) \Gamma_{a\,
\alpha\beta} \, , \qquad  \end{eqnarray} $g^{nm}(\xi)$ is inverse of the induced metric
$g_{mn}(\xi)= \hat{e}_m{}^a \hat{e}_{n\, a}$ (Eq. (\ref{ind-g}) with
$\hat{\theta}(\xi)=0$)
 and
\begin{eqnarray}\label{bgamT}
& {\bar{\gamma}}:= {i\over 3!\sqrt{|g(\xi) |}}\; \epsilon^{mnk}
  \hat{e}_m{}^{a} \hat{e}_n{}^{b}\hat{e}_k{}^{c}
 \Gamma_{abc} \;
\end{eqnarray}
({\it cf.} Eq. (\ref{bgam-SSP}) for $\hat{\theta}(\xi)=0$)  has  the properties
$\bar{\gamma}^2=I$, $tr(\bar{\gamma})=0$.
\item{{\bf 2)}} The  equations of motion for the bosonic supergravity
fields get (or may get) a source term contribution from the superbrane, while the
gauge--fixed equations for the bulk fermionic fields are sourceless (see Sec.2.3.1).

\item{{\bf 3)}} The action (\ref{SSG+M2}) possesses half of the
local supersymmetries of the pure supergravity action $S_{SG}[e^a, \psi^\alpha , A_3]$.
This is characterized \cite{BdAI01} by the standard transformation rules for the
supergravity fields \cite{CJS} (see Eqs. (\ref{susye}), (\ref{susyf-}), (\ref{susyA})
below) and by the following conditions (see Sec. 2.3.3) restricting the local
supersymmetry parameter on the worldvolume $W^3$,
\begin{eqnarray}
\label{susyKap} \hat{\varepsilon}^\alpha := \hat{\varepsilon}^\alpha(\hat{x}(\xi)) = (1
+ \bar{\gamma})^{\alpha}{}_{\beta}\kappa^\beta(\xi)\; .
\end{eqnarray}

\item{{\bf 4)}} The local supersymmetry algebra closes on-shell
in exactly the same manner as it does for the case of free supergravity (Sec. 2.3.4).

\end{itemize}

Property {\bf 1)} might seem strange since no worldvolume fermionic degrees of freedom
are seen directly in the gauge--fixed interacting action (\ref{SSG+M2}) involving the
{\it bosonic} brane action. But this {\it `fermionic equation for the bosonic brane'}
can be derived \cite{BdAI01,BAIL03} from the selfconsistency condition ${\cal
D}\Psi_{10\, \alpha} = 0$ for the gravitino equation $\Psi_{10\, \alpha} = 0$ (Eq.
(\ref{RS11}) below) which, according to  {\bf 2)}, remains sourceless in the presence
of the bosonic brane \cite{BdAI01}. Thus, it is convenient  to discuss property {\bf
2)} first.

\subsubsection{Field equations for the {\it SG--M2} system (property  {2})}

Varying the CJS action with respect to differentail forms $\delta S_{SG} = - 2i \int
\Psi_{10\, \alpha} \wedge \delta \psi^\alpha + \int {\cal G}_8 \wedge \delta A_3 + \int
M_{10 \, a} \wedge \delta E^a$ one can write the `free' supergravity equations in
differential form notation (see \cite{J+S99} and, {\it e.g.} \cite{BdAPV05}). The same
can be done for the (gauge--fixed) field equations of the {\it SG--M2} interacting
system, $\delta (S_{SG}+S_{M2}^0) = - 2i \int \Psi_{10\, \alpha} \wedge \delta
\psi^\alpha + \int ({\cal G}_8-J_8) \wedge \delta A_3 + \int (M_{10 \, a} - J_{10 \,
a})\wedge \delta E^a $. The variation of the bosonic membrane part $S_{M2}^0$ in the
action is written as an integral over spacetime $M^{11}$ with the use of the currents
(see \cite{BdAI01,BAIL03}; $\hat{e}^b:= d\xi^n \hat{e}_n{}^b(\xi)$, $\hat{e}_n{}^b:=
\partial_n\hat{x}^\mu(\xi) e_\mu {}^b  (\hat{x})$)
\begin{eqnarray}
\label{J10a} J_{10\, a}   =    {1\over 2 {e}(x) } e^{\wedge \, 10}_b \,
\int\limits_{W^3} \hat{\ast}\hat{e}_a \wedge \hat{e}^b   \delta^{11} (x -\hat{x}(\xi))
\; & =
 e^{\wedge \, 10}_b  \int\limits_{W^3} d^3\xi \, {\sqrt{|g|} \, g^{mn} \hat{e}_{n a}
\hat{e}_m^b  \over 2 |det(e_\nu{}^c(\hat{x}))| } \,  \delta^{11}(x -\hat{x}(\xi)) \, ,
\qquad
\\
\label{J8=} J_8   =  {1\over {e}(x) } e^{\wedge \, 8}_{abc} \int\limits_{W^3} \hat{e}^a
\wedge \hat{e}^b \wedge \hat{e}^c  \, \delta^{11} (x -\hat{x}(\xi)) & = e^{\wedge \,
8}_{abc} \int\limits_{W^3} d^3\xi {\epsilon^{mnk} \hat{e}_m^a \, \hat{e}_n^b\,
\hat{e}_k^c\over |det(e_\nu{}^c(\hat{x}))| }
 \, \delta^{11} (x
-\hat{x}(\xi))
  \, , \qquad {} \quad
\end{eqnarray}
which describe the brane source terms in the Einstein and gauge field equations. The
Einstein and the Rarita--Schwinger equations of the interacting system are written in
terms of the ten--forms
\begin{eqnarray}
\label{M10=J10} & M_{10\, a}  &:= {1\over 4} R^{bc} \wedge e^{\wedge 8}_{abc} + {1\over
2}\left( i_a F_4 \wedge \ast F_4 + F_4 \wedge i_a(\ast F_4) \right) + {\cal O}
(\psi^{\wedge 2}) + {\cal O} (\psi^{\wedge 4}) \; = J_{10\, a}\; ,
\\  \label{RS11}  & \Psi_{10\,
\alpha} &:= {\cal D} \psi^\beta \wedge \bar{\Gamma}^{(8)}_{\beta\alpha} = 0 \; ,
\end{eqnarray}
while the eight--form expression of the three--form gauge field equation reads
\begin{eqnarray}
\label{G8=J8}  & {\cal G}_8 &:=  d(\ast F_4 + b_7 -A_3\wedge dA_3) = J_8\; , \qquad
b_7:= {i\over 2} \psi^\alpha \wedge \psi^\beta   \wedge
\bar{\Gamma}^{(5)}_{\alpha\beta}  \; .
 \end{eqnarray}
In Eqs. (\ref{M10=J10}), (\ref{RS11}) and (\ref{G8=J8}) the eight--forms $e^{\wedge
8}_{abc}$, $\bar{\Gamma}^{(8)}_{\beta\alpha}$ and the five form
$\bar{\Gamma}^{(5)}_{\beta\alpha}$ are defined by
 \begin{eqnarray}\label{e8} & e^{\wedge (11-q)}_{a_1\ldots \, a_q}
 := {1\over (11-q)!} \epsilon_{a_{_1}\ldots\, a_qb_{_1}\ldots\, b_{_{11-q}}} e^{b_{_1}}
 \wedge \ldots \wedge e^{b_{_{11-q}}}\; , \qquad {\bar{\Gamma}}{}^{(q)}
 := {1\over q!} {e}^{a_q} \wedge \ldots \wedge {e}^{a_{_1}} \Gamma_{a_{_1}\ldots\, a_q}\; ,
\end{eqnarray}
the four--form $F_4$ is  the `supersymmetric' field strength of the three--form gauge
field $A_3$,
\begin{eqnarray}
\label{dA3=a+F} & F_4:= {1\over 4!} e^{a_4} \wedge \ldots \wedge e^{a_1} F_{a_1 \ldots
a_4}(x)  = dA_3 -  {1\over 2} \psi^\alpha\wedge \psi^\beta \wedge
\bar{\Gamma}^{(2)}_{\alpha\beta} \; ,  \qquad \;  \qquad
\end{eqnarray}
and $*F_4:= - e^{\wedge 7}_{abcd} F^{abcd}$. The spin connection
$\omega^{ab}=-\omega^{ba}$ are expressed through the graviton and gravitino by the
solution of the torsion constraint
\begin{eqnarray}\label{ta=}
T^a(x):= De^a = de^a - e^b\wedge \omega_b{}^a = -i \psi \wedge \Gamma^a\psi  \quad ,
\end{eqnarray}
which formally coincides with the leading component  of the on--shell superspace
constraint (\ref{cTa=}). The explicit expressions for the two--fermionic and
four--fermionic contributions, ${\cal O} (\psi^{\wedge 2})$ and ${\cal O} (\psi^{\wedge
4})$,  to the Einstein equations (\ref{M10=J10}) will not be needed in this paper. The
generalized covariant derivative ${\cal D}{\psi}^\alpha$ in (\ref{RS11}) is defined by
\begin{eqnarray}
\label{cD-def00} {\cal D}{\psi}^\alpha =
 d\psi^\alpha - {\psi}^\beta \wedge \omega_\beta{}^\alpha  \; , \quad
& \label{gom-def0}
 \omega_\alpha{}^\beta= {1\over 4}
\omega^{ab} \Gamma_{ab\alpha}{}^\beta + e^a t_{ a \alpha}{}^\beta
\end{eqnarray} and contains, in addition to the
spin connection ${1\over 4} \omega^{ab} \Gamma_{ab}{}_{\alpha}{}^\beta$, the covariant
contribution $e^a t_{a \alpha}{}^\beta $,
\begin{eqnarray}\label{t1-def0}
  & t_{a\, \beta}{}^\alpha :=  {i\over 18}  \left(F_{ac_1c_2c_3}
\Gamma^{c_1c_2c_3}{}_\beta^{\;\; \alpha} - {1\over 8} \, F^{c_1c_2c_3c_4}
\Gamma_{ac_1c_2c_3c_4}{}_\beta^{\;\; \alpha} \right) \; ,
\end{eqnarray}
expressed through the `supersymmetric' field strength $F_{abcd}(x)$ of $A_3$, Eq.
(\ref{dA3=a+F}). This covariant part of the generalized connection thus describes the
coupling of the bulk gravitino field to the fluxes of the three--form gauge field
$A_3$.

The reason for the absence of source in the fermionic equation (\ref{RS11}) obtained by
varying the gauge--fixed action (\ref{SSG+M2}) with respect to the gravitino field is,
clearly,  that the bosonic brane action $S^0_{M2}$ in (\ref{SSG+M2}) does not include
the gravitino $\hat{\psi}^\alpha$; this, in turn, follows from the absence of the
fermionic supervielbein ($E^\alpha(\hat{Z})$) in the supermembrane action
$\mathbf{S}_{M2}$ of Eq. (\ref{SM2ssp}). Nevertheless, the absence of an explicit
source term in (\ref{RS11}) does not imply that the gravitino is decoupled from the
brane source since Eq. (\ref{RS11}) includes the vielbein $e_\mu^a(x)$ (entering also
through the composite spin connection $\omega^{ab}$) and the field strength of the
three--form gauge field $A_3$ that do obey the sourceful Eqs. (\ref{M10=J10}) and
(\ref{G8=J8}).

Notice that the system of interacting equations, including Eqs. (\ref{M10=J10}),
(\ref{G8=J8}), (\ref{RS11}), (\ref{XIM2-c}) as well as the bosonic equation for the
brane, admits particular solutions with $\psi^\alpha(x)=0$. Inserting
$\psi^\alpha(x)=0$ back into the equations one arrives at the well--known system of
purely bosonic supergravity equations in \cite{Duff94}.

\subsubsection{`Fermionic equations' for the bosonic brane
interacting with supergravity (property 1)}

To understand how the `fermionic equation for the bosonic brane' results from the
consistency conditions of the gravitino equation one can use the identity (see {\it
e.g.} \cite{BdAPV05})
\begin{equation}\label{NIsusy}
{\cal D}\Psi_{10\,\alpha} =  i \psi^\beta \wedge \left( M_{10\, a}
\Gamma^a_{\beta\alpha} + {i\over 2} {\cal G}_8 \wedge \bar{\Gamma}^{(2)}_{\beta\alpha}
\right) \;
\end{equation}
that expresses the generalized covariant derivative of the {\it l.h.s.} of the
fermionic equation (\ref{RS11}) in terms of the left hand sides of the Einstein and the
$A_3$ gauge field equations, $M_{10\, a}$ and ${\cal G}_8$ in Eqs. (\ref{M10=J10}) and
(\ref{G8=J8}), respectively. For `free' $D=11$ supergravity the equations of motion are
$\Psi_\alpha =0$, $M_{10\, a}=0$ and ${\cal G}_8=0$, and the eleven--form {\it
identity} (\ref{NIsusy}) shows their interdependence. This {\it Noether identity}
reflects a local gauge symmetry of the CJS supergravity action $S_{SG}[e^a, \psi^\alpha
, A_3]$, the local supersymmetry of $D=11$ supergravity \cite{CJS} (Eqs.
(\ref{susye})--(\ref{susyA}) below).

When supergravity interacts with a bosonic membrane, $S_{SG}\mapsto S_{SG}  +S^0_{M2}$
like in the gauge fixed description, the bosonic field equations acquire source terms
and read $M_{10\, a}= J_{10\, a}$ [Eqs. (\ref{M10=J10}), (\ref{J10a})] and ${\cal G}_8=
J_8$ [Eqs. (\ref{G8=J8}), (\ref{J8=})]; the fermionic equation, however,  remains
sourceless, $\Psi_{10 \alpha}=0$ [Eq.  (\ref{RS11})]. Hence,  Eq. (\ref{NIsusy})
produces the following equation for the M2--brane currents
 $J_{10\, a}$ and $J_8$ (see  Eqs. (\ref{J10a}) and (\ref{J8=}))
\begin{equation}
\label{NIsusy-J}
  i \psi^\beta \wedge \left( J_{10\, a} \Gamma^a_{\beta\alpha}
  + {i\over 2} {J}_8 \wedge \bar{\Gamma}{}^{(2)}_{\beta\alpha}
\right) =0 \; .
\end{equation}
Due to the currents, this eleven--form equation has support on the M2--brane
worldvolume $W^3$ and so it can be written as a three--form equation  on $W^3$ in terms
of the pull--backs of the graviton and gravitino \cite{BdAI01,BAIL03}. When $S_{SG}
+S^0_{M2}$ provides the gauge fixed description of the supergravity--supermembrane
interaction the currents $J_{10\, a}$ and $J_8$ are defined by Eqs. (\ref{J10a}) and
(\ref{J8=}) and the equivalent form of Eq. (\ref{NIsusy-J}) reads
\begin{eqnarray}
\label{fermiM2} & \hat{\psi}^\beta \wedge\left( i \, \hat{\ast} \hat{e}^a  \,
\Gamma_{a\, \beta \alpha} + {1\over 2} \hat{e}^b\wedge \hat{e}^c\, {\Gamma}_{bc\, \beta
\alpha} \right)  = 0 \; .
\end{eqnarray}
A simple algebra allows us to present Eq. (\ref{fermiM2}) in the form of
\begin{eqnarray}
\label{XIM2}  \hat{\Xi}_{3\, \alpha} &:=&  \hat{\ast} \hat{e}^a \wedge \hat{\psi}^\beta
\, (\Gamma_a(1-\bar{\gamma}))_{\beta \alpha}  = 0 \;
\end{eqnarray}
where the action of $\hat{\ast}$ is defined by $\hat{\ast} \hat{e}^a = {1\over 2}
d\xi^{n} \wedge d\xi^m \epsilon_{mnk}\sqrt{|g(\xi)|}g^{kl}
\partial_l \hat{x}^\mu  e_\mu{}^{a}(\hat{x})$ (Eq.  (\ref{hHodge}) with
$\hat{\theta}=0$, {\it i.e.} for $\hat{\ast}= \hat{\star}\vert_{\hat{\theta}=0}$). Eq.
(\ref{XIM2}) is an equivalent form of Eq. (\ref{XIM2-c}) written in a conventional
differential form notation,
\begin{eqnarray}
\label{XIM2+c}  \hat{\Xi}_{3\, \alpha} &\propto & d^3 \xi\; \hat{\psi}_m
\hat{\Gamma}^m(1-\bar{\gamma}) \; .
\end{eqnarray}

\subsubsection{Supersymmetry of the gauge--fixed action (property 3)}

Eq. (\ref{NIsusy}) is the Noether identity for the local supersymmetry of the pure
supergravity action  $S_{SG}[e^a,\psi^\alpha, A_3]$,
\begin{eqnarray}
\label{susye} \delta_{\varepsilon}e^a &=& -2i {\psi} \Gamma^a{\varepsilon}:= - 2i
{\psi}^\alpha \Gamma^a_{\alpha\beta}{\varepsilon}^\beta \; ,
\\
\label{susyf-} \delta_{\varepsilon}\psi^\alpha &=& {\cal D}\varepsilon^\alpha(x) =
D\varepsilon^\alpha(x) - \varepsilon^\beta(x) e^a t_{a\beta}{}^\alpha(x) \; ,  \qquad
\\
\label{susyA} \delta_{\varepsilon}A_3 &=& \psi^\alpha \wedge
\bar{\Gamma}^{(2)}_{\alpha\beta}{\varepsilon}^\beta \; ,
\end{eqnarray}
where ${\cal D}$ is the generalized covariant derivative of Eq. (\ref{cD-def00}),
$D=d-\omega$ is the standard covariant derivative, and the tensorial part of the
generalized connection $t_{a\beta}{}^\alpha(x)$ is defined in Eq. (\ref{t1-def0}). The
fact that Eq. (\ref{NIsusy}) is not identically satisfied in the presence of a bosonic
brane, {\it i.e.} when $S_{SG}\mapsto S_{SG}+S_{M2}^0$, reflects the fact that the
bosonic brane action $S_{M2}^0$ breaks the local supersymmetry
(\ref{susye})--(\ref{susyA}).

 When the bosonic brane is the purely bosonic ($\hat{\theta}=0$)
`limit' of a superbrane,  the sum of the supergravity action and the action of bosonic
brane provides a gauge--fixed description of the supergravity---{\it super}brane ({\it
SG--M2}) interacting system \cite{BAIL03,BAIL-BI} and preserves one--half of the local
supersymmetry   of $S_{SG}$. This half of the local supersymmetry is defined by the
restriction (\ref{susyKap}) on the pull--back of the supersymmetry parameter to the
membrane worldvolume $W^3$, $\hat{\varepsilon}^\alpha :=
\hat{\varepsilon}^\alpha(\hat{x}(\xi)) = (1 +
\bar{\gamma})^{\alpha}{}_{\beta}\kappa^\beta(\xi)$. Its preservation  can be shown in
two ways, either explicitly \cite{BdAI01} (see also Sec. 3.3 below) or using the fact
that the action (\ref{SSG+M2}) provides a gauge--fixed version of the hypothetical
superfield description of the supergravity--superbrane interaction
\cite{BAIL-BI,BAIL03} as discussed also in Sec. 2.2.

\subsubsection{On--shell closure  of the local supersymmetry algebra
in the spacetime gauge--fixed description of {\it SG--M2} system (property 4)}

As known from the pioneering paper \cite{CJS}, the local supersymmetry transformations
(\ref{susye})--(\ref{susyA}) that leave invariant the supergravity action $S_{SG}[e^a,
\psi^\alpha, A_3(x)]$ form an algebra which is closed on shell, {\it i.e.} using  the
`free' supergravity equations.
 The structure  of  this algebra is schematically \cite{CJS}
\begin{eqnarray}\label{susyAl}
{}[\delta_{\epsilon_1} \, , \, \delta_{\epsilon_1}]\; [fields] = \left( \delta_{b^\mu
(\epsilon_1, \,
 \epsilon_2) } + \delta_{L^{ab}(\epsilon_1, \, \epsilon_2) } +
\delta_{\epsilon_3(\epsilon_1, \, \epsilon_2) } + {\cal K}_{(\epsilon_1, \,
\epsilon_2)} \right) \, [fields] \; ,
\end{eqnarray}
where $\delta_{b^\mu }$ determines the general coordinate transformations,
$\delta_{L^{ab} }$ the local Lorentz transformations, $\delta_{\epsilon}$ the local
supersymmetry transformations (\ref{susye})--(\ref{susyA}) and ${\cal K}_{(\epsilon_1,
\, \epsilon_2)}[fields]$ denotes terms that express the non--closure of the algebra and
that become zero on shell.

Let us now consider the {\it SG--M2} interacting system. The form of the supersymmetry
transformations leaving invariant the coupled supergravity--bosonic brane action
(\ref{SSG+M2}) ({\it i.e.} preserving the gauge $\hat{\theta}(\xi)=0$ for the
interacting supergravity--superbrane system \cite{BAIL-BI,BAIL03,BI03}) is exactly the
same as that of the supersymmetry of `free' supergravity \footnote{Notice that this is
not the case for the supersymmetric brane world models in \cite{Bagger+}. There, the
{\it brane} actions also contain the pull--back of the gravitino field. Probably these
two facts are related and prevent or hamper a superfield formulation of the brane
actions of \cite{Bagger+}. In all other respects the models of \cite{Bagger+} are
similar to the dynamical systems of supergravity interacting with standard superbranes
as they are presented in the gauge--fixed description of \cite{BAIL-BI,BdAI01} and Sec.
2 of this paper. The breaking of 1/2 of the supersymmetry in the gauge--fixed
description corresponds to imposing a kind of boundary conditions on  the supersymmetry
parameter in \cite{Bagger+}.}. However, in principle, the last term in (\ref{susyAl})
might spoil the closure of the local supersymmetry algebra of supergravity--bosonic
membrane system that provides the gauge--fixed description of the {\it SG--M2}
interacting system). This is not so, however.
 On the bosonic fields of the supergravity multiplet, $e_\mu{}^a(x)$ and
$A_{\mu\nu\rho}(x)$, the algebra (\ref{susyAl}) is closed off shell \cite{CJS}, {\it
i.e.} without any use of the equations of motion. This means that
\begin{eqnarray}\label{susy-eA}
{\cal K}_{(\epsilon_1, \, \epsilon_2) } [e^a(x)] \equiv 0 \; , \qquad {\cal
K}_{(\epsilon_1, \, \epsilon_2) } [A_3(x)] \equiv 0 \; .
\end{eqnarray}
Hence the on--shell character of the supersymmetry algebra comes from the
 fermionic fields since only  ${\cal K}_{(\epsilon_1, \, \epsilon_2) }[\psi]
\not=0\; $ off--shell \footnote{The statement that in the absence of the auxiliary
fields ${\cal K}_{(\epsilon_1, \, \epsilon_2) } [fermionic\; fields] \not=0$ off--shell
while ${\cal K}_{(\epsilon_1, \, \epsilon_2) } [bosonic\; fields]
 =0$  seems to be quite general, {\it i.e.} valid for many supersymmetric
theories in various  dimensions, see {\it e.g.} \cite{PvanN81}. To our  knowledge, the
only  exception is provided by the supersymmetry transformations that preserve the
equations of motion for supermultiplets that include self--dual gauge fields, where the
selfduality condition for the bosonic gauge field is also needed to close the
supersymmetry algebra.}. Moreover, and this is the key point, only the fermionic
equations are necessary to close supersymmetry algebra on the fermionic fields;
schematically,
\begin{eqnarray}\label{susy-eB}
{\cal K}_{(\epsilon_1, \, \epsilon_2) } [\psi^\alpha(x)] \; \propto \;  \ast \Psi_{10
\alpha} \; .
\end{eqnarray}
But as noticed above (following \cite{BdAI01,BAIL-BI,BAIL03}), the fermionic equation
for the interacting system in the gauge--fixed description given by the sum of
supergravity action and the action for bosonic brane preserving a half of the local
supersymmetry remains formally the same ({\it i.e.}, sourceless) as that for `free'
supergravity, $\Psi_{10 \alpha}=0$. Hence,
\begin{eqnarray}\label{susy-eA0}
{\cal K}_{(\epsilon_1, \, \epsilon_2) } [\psi^\alpha(x)] \vert_{_{{}^{on-shell \; for
\; the}_{interacting \; system}}} =0\; .
\end{eqnarray}
Further, the local supersymmetry transformations  act only on the fields of
supergravity multiplet, Eqs. (\ref{susye})--(\ref{susyA}), since the only supermembrane
field in the gauge--fixed description, the bosonic $\hat{x}(\xi)$, is inert under the
local spacetime supersymmetry. Thus the on-shell closure of the local supersymmetry
algebra of the gauge--fixed description of the supergravity--supermembrane interacting
system follows from that of the pure $D=11$ supergravity theory.

\setcounter{equation}0
\section{Goldstone nature of the supermembrane fermionic fields and
Dirac equation for the supermembrane in a D=11 supergravity background with fluxes}

The Goldstone nature of the superbrane coordinate functions, in particular of the
fermionic functions $\hat{\theta}^\alpha (\xi)$, has been known for a long time
\cite{H+L+Polchinski86,BST87}. For a superbrane interacting with {\it dynamical
supergravity} the $\hat{\theta}^\alpha (\xi)$  are Goldstone (or compensator) fields
for the local supersymmetry, a fact that explains the possibility of taking the gauge
(\ref{thGAUGE}), $\hat{\theta}(\xi)=0$,  by using this local supersymmetry (see Sec.
2.2).

In this gauge the Lagrangian description of the system is provided by the sum
(\ref{SSG+M2}) of the spacetime supergravity action without auxiliary fields and of the
bosonic M2--brane action \cite{BAIL-BI,BAIL03}. The full set of equations of motion is
given by the supergravity field equations (\ref{M10=J10}), (\ref{G8=J8}), (\ref{RS11}),
the bosonic brane equations ({\it cf.} Eq. (\ref{SEqmX}))
\begin{eqnarray}\label{EqmX}
D(\hat{\ast} \hat{e}_a)= -2 i_a F_4 - {1\over 2} \hat{e}^b \wedge \hat{\psi} \wedge
 \Gamma_{ab} \hat{\psi} \; , \qquad i_a F_4
 := {1\over 3!} \hat{e}^{d}\wedge\hat{e}^{c} \wedge \hat{e}^{b} \,
F_{abcd}(\hat{x})
  \; , \qquad \end{eqnarray}
and the `fermionic equation for  the bosonic brane' \cite{BdAI01}, Eq. (\ref{XIM2}),
({\it cf.} (\ref{SEqmTh}))
\begin{eqnarray}
\label{fermiM2-02} &&  \hat{\Xi}_{3\, \alpha}= \ast \hat{e}^a \wedge \hat{\psi}^\beta
\left( \Gamma_a(1-\bar{\gamma})\right)_{\beta \alpha}  = 0 \; . \quad
\end{eqnarray}
In the  $\hat{\theta}(\xi)=0$ gauge, the fermionic degrees of freedom of the
superbrane, usually associated with $\hat{\theta}(\xi)$, are contained in the
pull--back $\hat{\psi}^\beta$ of the bulk gravitino to the worldvolume $W^3$ as {\it
zero modes} corresponding to the supersymmetry broken by the brane \footnote{See
\cite{C+N=98} for a discussion of the brane degrees of freedom as zero modes, but
starting from certain brane solutions of the supergravity equations.}. Namely, the fact
that half of the supersymmetry is spontaneously broken by the presence of the
supermembrane is reflected by the explicit breaking of half of the local supersymmetry
by the bosonic brane in the gauge--fixed description (\ref{SSG+M2}). Hence, in the
presence of the supermembrane, the remaining local supersymmetry does not produce the
same number of gauge--fixing conditions on the gravitino field as the full local
supersymmetry of `free' supergravity. Nevertheless, as shown in \cite{BdAI01,BAIL-BI},
this super--Higgs effect in the presence of a superbrane does not make the gravitino
massive, because the `fermionic equation for the bosonic brane', Eq. (\ref{XIM2}),
takes the r\^ole of the lost gauge--fixing conditions and keeps the number of
polarizations of the gravitino equal to those in `free' supergravity. However, the
fermionic zero modes corresponding to the supersymmetry broken by the membrane remain
in the pull--back $\hat{\psi}$  of the bulk gravitino $\psi$ to $W^3$ \footnote{To
illustrate the above statement one can consider the weak field case, making a linear
approximation in the fields to find the general solution of the equations;
schematically, $\phi = \phi_0 + \sum (b(p) e^{+ip\cdot x}+ b^\dagger(p) e^{-ip\cdot
x})$. Then the number of polarizations (which distinguishes between the massive and
massless cases) is determined by the oscillating contributions, and depends on the
number of conditions imposed on the creation and annihilation operators $b^\dagger(p)$
and $b(p)$, while the zero modes are associated with the non--oscillating contribution
$\phi_0$.

The same occurs in the spacetime Higgs effect in general relativity interacting with
branes \cite{BdAIL-Higgs,BAIL-BI}. A bulk graviton does not get mass and keeps the same
number of polarizations in the presence of a $p$--brane because the bosonic equations
of the brane replace the gauge fixing conditions that were lost due to the spontaneous
breaking of the diffeomorphism symmetry by the $p$--brane
\cite{BdAI01,BAIL-BI,BdAIL-Higgs}. However, the corresponding gauge--fixing conditions
for the case of {\it free} gravity allow for a residual gauge symmetry, which is absent
when the bosonic equations of the brane take the r\^ole of the gauge fixing conditions.
This set of  residual gauge symmetries broken by the brane is the origin of the zero
modes of the graviton on $W^{p+1}$ that describe the brane degrees of freedom in the
`static' gauge (the statement in \cite{BdAI01,BAIL-BI,BdAIL-Higgs} that a $p$--brane
does not carry any local degrees of freedom in the presence of dynamical gravity refers
to the oscillating degrees of freedom -polarizations- of the graviton and gravitino, as
discussed above, not to the zero modes). We thank W. Siegel for an illuminating
discussion on this point.}. Precisely these zero modes represent the $16$ fermionic
degrees of freedom of supermembrane in the gauge--fixed description of (\ref{SSG+M2}).


 To summarize, in the gauge--fixed description of the
supergravity--supermembrane interaction provided by the set of equations
(\ref{M10=J10}), (\ref{G8=J8}), (\ref{RS11}), (\ref{EqmX}), (\ref{fermiM2-02}), the
bulk gravitino carries both the supergravity and the superbrane fermionic degrees of
freedom as determined by the solution of field equation (\ref{RS11}) with the boundary
conditions (\ref{fermiM2-02}) on the 3--dimensional `defect',  the brane worldvolume
$W^3$. This description is convenient in studying the cases where both the effects from
the bulk and from the worldvolume fermions are equally important and there is no need
to separate their contribution.

However, in some cases (interesting {\it e.g.}~for M-theory--based `realistic' model
building, see \cite{deW+PP98,HarveyMoore99,BranesSG,Dima+RK05,VP+2005}) it may happen
that the effects from the worldvolume fermions, and in particular the explicit form of
their interaction with the flux, constitute  the main interest. Then, when starting
from our gauge--fixed description, one faces the problem of visualizing the fermionic
degrees of freedom of the superbrane, {\it i.e.} the supermembrane coordinate functions
 $\hat{\theta}(\xi)$. This will be the main subject of the study below.

In the light of Goldstone nature of $\hat{\theta}(\xi)$, the general answer should not
be too surprising: the recovery  of the $\hat{\theta}(\xi)$ contributions to the action
and equations of motion can be done by making (consistently) a local supersymmetry
transformation the parameter of which  is identified with the Goldstone fermion field
$\hat{\theta}(\xi)$. We begin by showing how the supermembrane {\it fermionic equations
in a supergravity background with fluxes}, this is to say
 with nonvanishing  $F_{abcd}$, can be obtained on this way.

\subsection{Dirac equation for the  supermembrane in a
supergravity background with fluxes from the gauge--fixed approach}

When supergravity is treated as a background, one concentrates on the supermembrane
equations. In our  gauge--fixed description these are given by the bosonic equation
(\ref{EqmX}) and the fermionic Eq. (\ref{fermiM2-02}) which is more a condition on the
pull--back of the gravitino than  a dynamical equation. To separate the contribution
form the bulk fermions and from the supermembrane fermions one makes, following the
above prescription, the local supersymmetry transformations
(\ref{susye})--(\ref{susyA}) of the supergravity fields in (\ref{XIM2}) and identifies
the (pull--back of the) parameter of these transformations with the supermembrane
fermionic field, $\epsilon (\hat{x}(\xi))=\hat{\theta}(\xi)$. The result at first order
in $\hat{\theta}(\xi)$ is given by
\begin{eqnarray}
\label{fM2+susy} &&  \ast \hat{e}^a \wedge \hat{\psi}^\beta \left(
\Gamma_a(1-\bar{\gamma})\right)_{\beta \alpha} + \delta_{\hat{\epsilon}=\hat{\theta}}
\left( \ast \hat{e}^a \wedge \hat{\psi}^\beta \left(
\Gamma_a(1-\bar{\gamma})\right)_{\beta \alpha} \right)   = 0 \; , \quad
\end{eqnarray}
or, in more detail ($\hat{e}_m^a= \partial_m \hat{x}^\mu e_\mu^a(\hat {x})$,
$\hat{\psi}_m^\alpha= \partial_m \hat{x}^\mu {\psi}_\mu^\alpha(\hat {x})$,
$\hat{\Gamma}^k =g^{km}(\xi)\hat{e}_m^a \Gamma_a$, $g_{mn}(\xi)=\hat{e}_m^a
\hat{e}_{na} $),
\begin{eqnarray}\label{fM2+susy1}
\hat{\ast} \hat{e}^a &\!\! \wedge \left(\hat{\psi}{\Gamma}_a(1-\bar{\gamma}) + {\cal
D}\hat{\theta}{\Gamma}_a(1-\bar{\gamma})  + 2i \hat{\psi}_k\hat{\Gamma}^k\hat{\theta}\,
\hat{\psi}{\Gamma}_a  +   { \epsilon^{mnk} \hat{e}_m^{b_1}\hat{e}_n^{b_2}\over
\sqrt{|{g}(\xi)|}} (\hat{\psi}_k{\Gamma}^{b_3}\hat{\theta})\,
\hat{\psi}{\Gamma}_a{\Gamma}_{b_1b_2b_3} \right)_{\alpha} + \nonumber
\\
& + 2i \hat{\ast}\hat{e}^b \, \wedge (\hat{\psi} {\Gamma}_a(1-\bar{\gamma}))_{\alpha}
\; (\hat{\psi}^l{\Gamma}_b\hat{\theta} + {\psi}_b(\hat{x})\hat{\Gamma}^l\hat{\theta})\,
\hat{e}_l{}^a - 2i \hat{\ast}\hat{\psi}{\Gamma}^a\hat{\theta} \wedge (\hat{\psi}
{\Gamma}_a(1-\bar{\gamma}))_{\alpha}
  =0 \; ,
\end{eqnarray}
where again the generalized covariant derivative ${\cal D}$ is given  by Eqs.
(\ref{cD-def00}), (\ref{t1-def0}) and, thus, includes a  contribution from the fluxes
$F_{abcd}$. We have checked explicitly that Eq. (\ref{fM2+susy1}) {\it formally}
coincides with the first order equation that can be obtained within the standard
`background superfield' approach \cite{deW+PP98,HarveyMoore99,BranesSG,Dima+RK05}
(without setting $\hat{\psi}=0$ as in
\cite{deW+PP98,HarveyMoore99,BranesSG,Dima+RK05}). By `formally' we mean that in the
equations obtained in the standard framework the graviton, the gravitino and the gauge
field strength are, strictly speaking, solutions of the `free' supergravity equations,
while in our case such a restriction is absent and one can use, {\it e.g.}, solutions
of the interacting system of equations.

Eq. (\ref{fM2+susy1}) is rather complicated.   A simpler one results  when in
(\ref{fM2+susy1}) the gravitino field is set equal to zero. This gives
\begin{eqnarray}
\label{fM2+susy2}  && \hat{\psi}^\alpha  = 0 \; : \qquad  \hat{\ast} \hat{e}_a \wedge
{\cal D}\hat{\theta}^\beta  \left( \Gamma^a(1-\bar{\gamma})\right)_{\beta \alpha}  = 0
\; , \qquad
\end{eqnarray}
or, equivalently, $${\cal D}\hat{\theta}^\beta \wedge \big( i \; \hat{\ast} \hat{e}_a
 \Gamma^a_{\beta \alpha} +
\hat{\bar{\Gamma}}{}^{(2)}_{\beta \alpha}\big) = 0\; . $$ Eq. (\ref{fM2+susy2})
formally coincides with the M2--brane Dirac equation which is obtained in
\cite{deW+PP98,HarveyMoore99,Dima+RK05} within the on--shell background superfield
approach, namely by expanding  Eq. (\ref{SEqmTh}) in $\hat{\theta}$ for $\psi=0$. To
see this explicitly, one may use the expression (\ref{cD-def00}), (\ref{t1-def0}) for
the generalized covariant derivative ${\cal D}$ in (\ref{fM2+susy2}) and the
worldvolume tensor notation ($\hat{\Gamma}_n:= \hat{e}_n^a\Gamma_a$, $\hat{\Gamma}^n:=
g^{nm}(\xi) \hat{e}_m^a\Gamma_a$ {\it etc.}) to arrive at
\begin{eqnarray}
\label{fM2+susy5} &  \left(D_n\hat{\theta} + {i\over 18} \hat{e}_n^a
\left(F_{ab_1b_2b_3} \; \hat{\theta} \Gamma^{b_1b_2b_3}- {1\over 8} F^{b_1b_2b_3b_4}\;
\hat{\theta} \Gamma_{ab_1b_2b_3b_4}\right)
 \right)^\beta\, \left(
\hat{\Gamma}^n(1-\bar{\gamma})\right)_{\beta \alpha}  = 0 \; , \\ \nonumber &
D_n\hat{\theta} ^\alpha :=
\partial_n \hat{\theta}^\alpha(\xi) - {1\over 4} \hat{e}_n^c\omega_c^{ab} \,
\hat{\theta}^\beta (\xi) \Gamma_{ab}{}_\beta{}^\alpha \; .
\end{eqnarray}
In this form the interaction of the supermembrane fermionic field with the $A_3$
`fluxes', this is to say with  the  field strength $F_{abcd}$, is manifest.

Linearizing Eq. (\ref{fM2+susy1}) in {\it all} the fermions, {\it i.e.} ignoring ${\cal
O}(\hat{\theta} \,\hat{\psi})$, ${\cal O}(\hat{\psi}^{\wedge 2})$ together with the
${\cal O}(\hat{\theta}^{\wedge 2})$ contributions, we find the equation
\begin{eqnarray}\label{fM2+susy1-2}
\hat{\ast} \hat{e}^a &\!\! \wedge ({\cal D}\hat{\theta}+
\hat{\psi}){\Gamma}_a(1-\bar{\gamma})
  =0 \;
\end{eqnarray}
which includes the pull--back of the gravitino $\hat{\psi}$ and the Goldstone fermion
$\hat{\theta}$ in the combination $({\cal D}\hat{\theta}+ \hat{\psi})$ only, which is
invariant under the linearized supersymmetry. This observation supports the discussed
fact (see footnote 6 and above) that, in the gauge $\hat{\theta}=0$, the zero modes
describing the brane fermionic degrees of freedom appear in the pull--back $\hat{\psi}$
of the bulk gravitino to $W^3$.

\subsection{On the contribution of the supermembrane fermionic
field to the full set  of interacting equations}

{} From the point of view of the interacting system, the setting $\hat{\psi}=0$ above
(and in \cite{Dima+RK05,VP+2005,deW+PP98,HarveyMoore99,BranesSG}) or, taking into
account the previous supersymmetry transformations that make manifest the Goldstone
degrees of freedom, $\hat{\psi}^\alpha := {\psi}^\alpha(\hat{x}(\xi)) = {\cal
D}\hat{\theta}^\alpha(\xi)$ is a kind of ansatz, or boundary condition, for the
gravitino field on $W^3$. As such, its consistency with the supergravity equations
should be checked. This is a convenient point to begin discussing the contribution of
the supermembrane fermionic fields to the complete system of interacting equations,
which includes the field equations whose gauge--fixed form is given by Eqs.
(\ref{M10=J10}), (\ref{G8=J8}) and (\ref{RS11}).

It is natural to consider the above relation $\hat{\psi}^\alpha = {\cal
D}\hat{\theta}^\alpha(\xi)$ on $W^3$ as  produced by the ansatz
\begin{eqnarray}
\label{psi=Dt} {\psi}^\alpha(x) = {\cal D}\tilde{\theta}^\alpha(x)\; \qquad
\end{eqnarray}
for the bulk gravitino, where the tilde denotes function on spacetime. Here the
defining property of the Volkov--Akulov Goldstone fermion (see \cite{VA72,VS73})
$\tilde{\theta}^\alpha(x)$ is that  its pull--back on $W^3$ coincides with the
supermembrane fermionic field,
\begin{eqnarray}
\label{tth=ht} \tilde{\theta}^\alpha(\hat{x}(\xi))= \hat{\theta}^\alpha(\xi )\; .
\qquad
\end{eqnarray}
The irrelevance of the properties of $\tilde{\theta}^\alpha(x)$ outside the brane
worldvolume $W^3$ is just the statement of the local supersymmetry of the `free'
supergravity action.

However, a direct substitution of  the ansatz (\ref{psi=Dt}), (\ref{tth=ht}) into the
gravitino equations (\ref{RS11}) would produce a problem. After some algebra ({\it
e.g.} using identities from \cite{BdAPV05} and (\ref{NIsusy})) one finds that such a
Volkov--Akulov Goldstone fermion $\tilde{\theta}^\alpha(x)$ would obey $i
\tilde{\theta}^\beta  ( J_{10\, a} \Gamma^a_{\beta\alpha} + {i\over 2} {J}_8 \wedge
\bar{\Gamma}{}^{(2)}_{\beta\alpha}) =0$. This is equivalent ({\it cf.}  Sec. 2.3.2) to
the condition $\hat{\theta}^\beta (\xi) ( i\, \hat{\ast} \hat{e}^a  \, \Gamma_{a\,
\beta \alpha} + {1\over 2} \hat{e}^b\wedge \hat{e}^c\, {\Gamma}_{bc\, \beta \alpha}) =
0$ or, equivalently, $\hat{\theta}^\beta (\xi)
(\hat{\Gamma}_n(1-\bar{\gamma}))_{\beta\alpha}=0$ which implies the effective vanishing
of the supermembrane fermionic field (actually $(1-\bar{\gamma})\hat{\theta}=0$, but
this in turn implies $\hat{\theta}=0$, since the $(1+\bar{\gamma})\hat{\theta}$ part
can be removed by the preserved supersymmetry gauge transformations which correspond to
the $\kappa$--symmetry of the superbrane).

The reason for this apparent problem lies  in the fact that the correct prescription to
recover the supermembrane fermionic fields is to make the supersymmetry transformations
of the gauge--fixed equations rather than using an ansatz like (\ref{psi=Dt}),
(\ref{tth=ht}) in them. Despite that the {\it r.h.s.} of (\ref{psi=Dt}) coincides with
the gravitino supersymmetry transformations, its substitution into (\ref{RS11}) does
not automatically give the supersymmetry transformations of this equations. The point
is that in a gauge--fixed equation where some Goldstone fields are set equal to zero,
{\it e.g.} $\hat{\theta}^\alpha(\xi)=0$, a zero in the {\it r.h.s.} of this equation
may come from a term proportional to $\hat{\theta}^\alpha(\xi)$. As it is suggested by
the study of the superfield description of the $D=4$, $N=1$ supergravity--superparticle
and supergravity--superstring systems \cite{BAIL03,BI03}, this is exactly the case for
the gauge fixed form of the gravitino equation (\ref{RS11}). Namely, the fully
supersymmetric (not gauge--fixed) counterpart of this equation contains a {\it r.h.s.}
proportional to $\hat{\theta}^\alpha(\xi)$. Schematically,
\begin{eqnarray}
 \label{RS11=O(t)0}  & \Psi_{10\, \alpha} &:= {\cal D} \psi^\beta \wedge
\bar{\Gamma}^{(8)}_{\beta\alpha} = {\cal O}(\hat{\theta}^\alpha(\xi)) \; . \qquad
\end{eqnarray}
In other words,  $\Psi_{10\, \alpha} \propto \hat{\theta}^\alpha $ rather than zero
like in Eq. (\ref{RS11}) which comes from (\ref{RS11=O(t)0}) in the gauge
$\hat{\theta}=0$. Then, taking into account the presence of a right hand side
proportional to $\hat{\theta}^\alpha$ in a fully supersymmetric (not gauge--fixed)
counterpart of Eq. (\ref{RS11}), one can use a local supersymmetry transformation to
find an approximate expression for this  {\it r.h.s.} (${\cal O}(\hat{\theta})$ in
(\ref{RS11=O(t)0})) up to the first  order in $\hat{\theta}^\alpha$. This suggests a
way of deriving the contributions of the supermembrane Goldstone fermion
$\hat{\theta}^\alpha(\xi)$ to the supergravity equations from the local supersymmetry
transformations of the gauge--fixed system of interacting equations (\ref{M10=J10}),
(\ref{RS11}), (\ref{G8=J8}) or of the gauge--fixed interacting action (\ref{SSG+M2}),
which should work at least in low orders in $\hat{\theta}^\alpha(\xi)$.

As a first step in this direction let us derive the gravitino vertex operator of
\cite{BBS95} and, thus, find the contribution proportional to $ \hat{\theta}^\alpha $
in the right hand side of the fermionic field equation (\ref{RS11=O(t)0}).

\subsection{Gravitino vertex operator, a simple derivation of
the `fermionic equation for bosonic brane' and the Dirac action for the supermembrane
fermionic field}

The supersymmetry variation of the supergravity fields in the bosonic membrane action
gives
\begin{eqnarray}
\label{vsusySM2} \delta_{ \varepsilon} S^0_{M2}&=& {1\over 2}
 \int_{W^3} \hat{\ast} \hat{e}_a \wedge \delta_{ \varepsilon}e^a -
\int_{W^3} \delta_{ \varepsilon}\hat{A}_3  = - i \int_{W^3} \hat{\ast} \hat{e}_a \wedge
\hat{\psi}^\beta  \left( \Gamma^a(1-\bar{\gamma})\right)_{\beta
\alpha}\varepsilon^\alpha (\hat{x}) \; ,  \qquad
\end{eqnarray}
Notice that the requirement that this variation
 is zero for an arbitrary value of the
parameter $\varepsilon^\alpha (\hat{x}(\xi))$,
 $\delta_{ \varepsilon} S_{M2}=0$
 results in
\begin{eqnarray} \label{vs-fEq}
 \delta_{ \varepsilon} S_{M2}&=&0 \quad  \forall \varepsilon^\alpha (\hat{x}) \qquad
 \Rightarrow \qquad
  \hat{\ast}\hat{e}_a \wedge \hat{\psi}^\beta  \left(
\Gamma^a(1-\bar{\gamma})\right)_{\beta \alpha}= 0 \; ,
\end{eqnarray}
which is exactly the `fermionic equation for the bosonic brane', Eq. (\ref{XIM2}). This
easy way to derive Eq. (\ref{XIM2}) is, actually, equivalent to a more involved
derivation through the consistency conditions for the bulk field equations (see Sec.
2.3.2 and \cite{BdAI01}). As a byproduct one also easily sees that a supersymmetry
transformation with parameter restricted by (\ref{susyKap}), $\hat{\varepsilon}^\alpha
:= \hat{\varepsilon}^\alpha(\hat{x}(\xi)) = (1 +
\bar{\gamma})^{\alpha}{}_{\beta}\kappa^\beta(\xi)$, leaves the action $S_{M2}^0$ and
hence $S_{SG}+ S_{M2}^0$ invariant,
\begin{eqnarray}
\label{susyM2} \delta_{\varepsilon}S_{M2}^0\vert_{\hat{\varepsilon} = (1 +
\bar{\gamma})\kappa(\xi)}=0\qquad \Rightarrow \qquad \delta_{\varepsilon}(S_{SG}[e^a,
\psi^\alpha, A_3] + S_{M2}[\hat{e}^a, \hat{A}_3])\vert_{\hat{\varepsilon} = (1 +
\bar{\gamma})\kappa(\xi)}=0\; ,
\end{eqnarray}
which follows from the fact that $(1 - \bar{\gamma})(1 + \bar{\gamma})=0$. This
preserved supersymmetry, coming from  the $\kappa$--symmetry of the superbrane, allows
one to extend the identification of broken supersymmetry with physical fermionic
degrees of freedom of supermembrane,
$(1-\bar{\gamma})\hat{\varepsilon}=(1-\bar{\gamma})\hat{\theta}$ (as in
\cite{H+L+Polchinski86}), to a full  identification of the pull--back to $W^3$ of
supersymmetry parameter with the fermionic field,
\begin{eqnarray}
\label{hep=hth} &&  \hat{\varepsilon}^\alpha = \varepsilon^\alpha (\hat{x}(\xi))=
\hat{\theta}^\alpha(\xi) \; .
\end{eqnarray}
 With such an identification the expression for the interacting action becomes
\begin{eqnarray}\label{SSG+M2+}
S_{SG-M2}= S_{SG}[e^a , \psi^\alpha , A_3] + S_{M2}= S_{SG}[e^a , \psi^\alpha , A_3] +
S^0_{M2}[\hat{e}^a, \hat{A}_3] -i  \int_{W^3} {\Xi}_{3\alpha} \, \hat{\theta}^\alpha +
{\cal O}(\hat{\theta}^2) \; , \qquad
\end{eqnarray}
where the first two terms in the {\it l.h.s.}  describe the gauge--fixed action
(\ref{SSG+M2}), while the third term ({\it cf.} (\ref{XIM2})),
\begin{eqnarray} \label{K1=}
\int_{W^3} {\Xi}_{1\alpha}(\xi)  \hat{\theta}^\alpha := i \delta_{\hat{\varepsilon}=
\hat{\theta}} S_{M2} \; , \qquad {\Xi}_{3\alpha}(\xi) := \hat{\ast}\hat{e}_a \wedge
\hat{\psi}^\beta  \left( \Gamma^a(1-\bar{\gamma})\right)_{\beta \alpha} \; ,
\end{eqnarray}
is given by the supersymmetry variation of $S_{M2}^0$ by substituting $\hat{\theta}$
for $\hat{\varepsilon}$. This first order contribution determines the supermembrane
fermionic vertex operator ${\cal V}$ as defined in \cite{BBS95},
\begin{eqnarray}
\label{vth=Vpsi} && i \delta_{\hat{\varepsilon}= \hat{\theta}} S_{M2}=  \int_{W^3}
{\Xi}_{3\alpha}(\xi)  \hat{\theta}^\alpha = \int_{W^3} d^3 \xi \,
\psi_\mu{}^\beta(\hat{x}) \, {\cal V}_\beta{}^\mu (\xi) \; , \\ \label{V=}&&  d^3 \xi
 \,  {\cal V}_\beta{}^\mu (\xi) = d\hat{x}^\mu(\xi) \wedge  \hat{\ast} \hat{e}_a
 \, \left(\Gamma^a(1-\bar{\gamma}) \hat{\theta}(\xi)\right)_\beta
 =  d\hat{x}^\mu \wedge \left( \hat{\ast} \hat{e}_a
 (\Gamma^a \hat{\theta})_\beta  - i (\hat{\bar{\Gamma}}{}^{(2)} \hat{\theta})_\beta \right)
 \; . \quad
\end{eqnarray}
Thus, starting from a full but gauge--fixed description of the supergravity--superbrane
interaction of \cite{BAIL03,BAIL-BI}, we reproduce the supermembrane vertex operator
from \cite{BBS95}.

Notice that calculations as those above may apply  equally well to the action
(\ref{SSG+M2}) of the interacting system and to the action $S_{M2}$ of a bosonic
membrane in a spacetime supergravity background. Of course in the latter case the local
supersymmetry is not a gauge transformation of the action but rather a transformation
of the background fields.

By construction, the action (\ref{SSG+M2+}) is invariant under full local supersymmetry
(not just one--half as the gauge--fixed action (\ref{SSG+M2})) up to contributions
proportional to $\hat{\theta}$. Indeed, the Goldstone nature of $\hat{\theta}$ implies
$\delta_{\varepsilon}\hat{\theta}(\xi) = - {\varepsilon}(\hat{x}) + {\cal O}
(\hat{\theta})$ which, in the light of (\ref{vth=Vpsi}) and of the supersymmetry
invariance of $S_{SG}$, gives $\delta_{\hat{\varepsilon}} (S_{SG} + S_{M2})=
\delta_{\hat{\varepsilon}} (S_{SG} + S^0_{M2} - i\int \, \Xi_3 \hat{\theta}) = {\cal
O}(\hat{\theta})$ for the action (\ref{SSG+M2+}). To reach the supersymmetry invariance
up to the first order in $\hat{\theta}$ one needs to recover the ${\cal
O}(\hat{\theta}^{\wedge 2})$ components in the action. In our approach this can be done
by adding ${1\over 2} \delta_{\hat{\varepsilon}= \hat{\theta}} (S_{SG} + S^0_{M2} -
i\int \, \Xi_3 \hat{\theta})$ to the action (\ref{SSG+M2+}). This is just the term that
should produce the supermembrane fermionic equation (\ref{fM2+susy1}). One easily
checks this for $\psi=0$. Indeed,
\begin{eqnarray}\label{vSSG+M2+}
{1\over 2} \delta_{\hat{\varepsilon}= \hat{\theta}} \left(S^0_{M2} - i\int \, \Xi_3
\hat{\theta}\right)_{\vert \hat{\psi}=0}  = - {i\over 2} \int \,
\delta_{\hat{\varepsilon}= \hat{\theta}}\left(\Xi_3\right)_{\vert \hat{\psi}=0} \,
\hat{\theta} = -{i\over 2} \int \hat{\ast}\hat{e}_a \wedge {\cal D}{\hat{\theta}}
\Gamma^a(1-\bar{\gamma})\hat{\theta} \;
\end{eqnarray}
produces the Dirac equation (\ref{fM2+susy2}).

The action (\ref{SSG+M2+}), linear in $\hat{\theta}$, allows us to derive a
supersymmetric set of interacting equations for the supergravity--supermembrane system
with the same accuracy. For instance, the supersymmetric gravitino equation reads ({\it
cf. } (\ref{RS11}); notice that $\Gamma^a(1-\bar{\gamma})\hat{\theta}=
\hat{\theta}(1+\bar{\gamma})\Gamma^a= \hat{\theta}\Gamma^a(1-\bar{\gamma})$)
\begin{eqnarray}
 \label{RS11=O(t)}  \Psi_{10\, \alpha} &:= & {\cal D} \psi^\beta \wedge
\bar{\Gamma}^{(8)}_{\beta\alpha} = J_{10\, \alpha}[\hat{\theta}]  + {\cal
O}(\hat{\theta}^2) \; ,
\\ \label{Jf10=}
&  J_{10\, \alpha}[\hat{\theta}] & =  {i\over 2e(x)} e_b^{\wedge 10} \int_{W^3}
\hat{\ast} \hat{e}_a \wedge \hat{e}^b \left( \hat{\theta}(\xi)\Gamma^a(1-\bar{\gamma})
\right)_\alpha\; \delta^{11} (x-\hat{x}(\xi)) \;  .
\end{eqnarray}

 Now, removing the bulk fermion by inserting the ansatz
(\ref{psi=Dt}) for $\psi(x)$ in (\ref{RS11=O(t)}) and ignoring higher order terms in
$\hat{\theta}$, one finds the relation between the bosonic currents (\ref{J10a}),
(\ref{J8=}) and the
 fermionic current (\ref{Jf10=}),
\begin{eqnarray}
 \label{tthJ=Jf}
\tilde{\theta}(x)\left(iJ_{10\, a} \Gamma^a_{\beta\alpha} - {1\over 2} {J}_8 \wedge
\bar{\Gamma}{}^{(2)}_{\beta\alpha}\right)=
 J_{10\, \alpha}[\hat{\theta}]  \; ,
\end{eqnarray}
which is satisfied identically for a Goldstone fermion obeying (\ref{tth=ht}). This
shows that it is consistent to use the ansatz (\ref{psi=Dt}) to study particular
solutions for the interacting system of supergravity and superbrane. Although this
consistency is widely believed, the above is, to our knowledge, its first explicit
check within the fully interacting system.

The study of the first order contribution in $\hat{\theta}$ to the full  system of
interacting equations for the $D$=11 supergravity--supermembrane system, as well as for
systems including M5--brane and $D$=10 Dirichlet superbranes is a problem for further
study. Another interesting question is whether one can extend the present approach to
include contributions  of higher order in $\hat{\theta}$ by using a counterpart of
Noether method (see \cite{PvanN81}) or, better still, the gauge completion procedure
(see \cite{CF80,deW+PP98}) but applied to the action as a whole rather than to the
construction of the supervielbein and other separate superfields.

\setcounter{equation}0
\section{Conclusions and discussion}

In this paper we have shown how the Dirac equation (\ref{fM2+susy2}) [(\ref{fM2+susy1})
for $\psi\not=0$] for the fermionic coordinate field $\hat{\theta}(\xi)$  of the
supermembrane (see \cite{deW+PP98,HarveyMoore99,BranesSG,Dima+RK05}) can be reproduced
from a complete but gauge--fixed Lagrangian description of the D=11
supergravity--supermembrane interacting system \cite{BAIL03,BAIL-BI}. This component
spacetime Lagrangian description is provided by the sum of the Cremmer--Julia--Scherk
supergravity action \cite{CJS} and a bosonic brane action given by the purely bosonic
($\hat{\theta}=0$) `limit' of the supermembrane action \cite{BST87}. It preserves half
of the local supersymmetry \cite{BdAI01} reflecting  the $\kappa$--symmetry of the
superbrane action. From the point of view of the hypothetical superfield action for the
supergravity--supermembrane interacting system the above spacetime description appears
\cite{BAIL03,BAIL-BI,BI03} as a result of fixing the superdiffeomorphism and superspace
Lorentz symmetry by choosing the Wess--Zumino gauge for the supergravity superfields
and of fixing (half of) the local supersymmetry by the $\hat{\theta}^\alpha(\xi)=0$
gauge for the superbrane.

Formulated as a general prescription, our way of deriving the superbrane equations of
motion consists in performing a  spacetime local supersymmetry transformation
[$\delta_{\varepsilon}$ of Eqs. (\ref{susye})--(\ref{susyA})] on the component fields
that appear in the `fermionic equation for bosonic brane' \cite{BdAI01}
[$\hat{\Xi}_{3\alpha}=0$, Eq. (\ref{XIM2})], and then identifying the (pull--back of
the) parameter of this transformation with the superbrane fermionic field
$\hat{\theta}(\xi)$ [thus $\hat{\Xi}_{3\alpha}+ \delta_{\hat{\varepsilon}=
\hat{\theta}} \hat{\Xi}_{3\alpha}=0$]. The identification of the $\hat{\theta}(\xi)$
with the parameter of the supersymmetry ($\hat{\varepsilon}= \hat{\theta}$) is made
possible by the Goldstone nature of this superbrane fermionic field: its (non--pure
gauge with respect to the $\kappa$--symmetry) components are the Goldstone fermions for
the supersymmetries spontaneously broken by the superbrane \cite{H+L+Polchinski86}.

 The original `fermionic
equation for the bosonic brane' ( $\hat{\Xi}_{3\alpha}=0$, Eq. (\ref{XIM2})) is
obtained as a consistency condition for the bosonic and fermionic
 field equations of the gauge fixed description of the supergravity--superbrane
interacting system \cite{BAIL03,BAIL-BI,BI03} which {\it does not} involve the
superbrane fermionic $\hat{\theta}^\alpha(\xi)$ variable explicitly\footnote{Our
approach  makes particularly clear why the Dirac equation for the superbrane in a
supergravity background with $\hat{\psi}=0$ contains the same generalized covariant
derivative (${\cal D}=D-t= d-\omega -t$) involved in the gravitino supersymmetry
transformation rules, a point also emphasized in Sec. 3 of a recent paper
\cite{DS+EB+RK}, where our ${\cal D}\hat{\theta}$ is denoted by $\delta \psi \,
\theta$. In the standard on-shell superfield approach such a coincidence can be traced
to the fact that the component supersymmetry transformations may be deduced from the
on--shell superspace constraints of \cite{CF80,BH80}.}. Here, in Sec. 3.1, we have also
shown how this `fermionic equation for the bosonic brane' (\ref{XIM2}) can be obtained
in an equivalent but very simple way, using as above the local supersymmetry
transformation with $\hat{\varepsilon}= \hat{\theta}$, but for the bosonic brane
action.  In this way one also recovers the gravitino vertex operator of \cite{BBS95}.
One may also notice that the `fermionic equation for bosonic brane' formally coincides
with the result of setting $\hat{\theta}^\alpha(\xi)=0$ in the most general form of the
superfield fermionic equations for superbranes in an on--shell superfield supergravity
background, Eqs. (\ref{SEqmTh}). Namely the leading component of (\ref{SEqmTh}) gives
(\ref{XIM2}) but with the graviton and the gravitino satisfying the `free' supergravity
equations of motion, which is not the case for Eq. (\ref{XIM2}) derived from the
complete spacetime Lagrangian description. Moreover, this situation holds at least at
first order in $\hat{\theta}$ for the fermionic equations of motion and at second order
in $\hat{\theta}$ for the action, namely our equation for the supermembrane Goldstone
fermion $\hat{\theta}(\xi)$ also coincides (formally) with the equations derived in
\cite{deW+PP98,HarveyMoore99,Dima+RK05}.

This shows, as widely believed, that the linearized equation for $\hat{\theta}(\xi)$
derived from the standard on--shell superfield approach to the supergravity background
is still valid for the case of background fields that are not restricted by the `free'
supergravity equations, in spite of the fact that the on--shell constraints implying
these `free' supergravity equations were an essential ingredient in the derivation of
the Dirac equation within the usual background on-shell  superfield approach. Notice
that our results also fit with those of \cite{deW+PP98} where it was found that,
although the complete $\kappa$--symmetry of the supermembrane action (\ref{SM2ssp}) in
curved superspace requires that the supervielbein $E_M^A(Z)$ and the super-3-form
$A_3(Z)$ obey the on--shell supergravity constraints, the requirement of
$\kappa$--symmetry up to the first order in $\hat{\theta}$ for the action written up to
the second order in $\hat{\theta}$ does not impose any restrictions on the component
background fields. Namely \cite{deW+PP98}, if the on-shell supergravity constraints are
used to decompose the action (\ref{SM2ssp}) in powers of $\hat{\theta}$ neglecting
${\cal O}(\hat{\theta}^{\wedge 3})$ terms and, then, the $\kappa$--symmetry is checked
neglecting ${\cal O}(\hat{\theta}^{\wedge 2})$ terms, the result is that, surprisingly,
such a weakened $\kappa$--symmetry requirement does not restrict the background fields
of the supergravity multiplet by any equations of motion. An important question is
whether this is also the case for the decomposition of the standard supermembrane
action including higher order ${\cal O}(\hat{\theta}^{\wedge 3})$  terms in
$\hat{\theta}$, and, if so, whether such a decomposition would coincide with the action
obtained by a development of the approach of the present paper.

Within the  on--shell background  superfield approach such calculations, also
technically involved, are possible using the recent results of \cite{Tsimpis04}. To
obtain equations of motion with  higher order $\hat{\theta}(\xi)$ terms in present
approach one has to perform a {\it `non--infinitesimal' supersymmetry transformation}
up to some power in the parameter; the finite supersymmetry transformation, if found,
might produce the fully supersymmetric (not gauge--fixed) action, if exists. For the
existence of such finite transformation it is important that the local supersymmetry of
the component gauge fixed description of the supergravity--supermembrane system is
closed at least on shell. We have shown in Sec. 2.2.4 that this is indeed the case and
that this follows from the closure of the local supersymmetry of free
supergravity\footnote{It would be interesting to study the algebra of the spacetime
local supersymmetry of the $D=11$ supergravity interacting with M5--brane and of the
$D=10$ supergravity interacting with higher Dirichlet branes. The (spacetime,
gauge--fixed) Lagrangian description of such interactions implies the use of the
duality--invariant formulations of supergravity (see \cite{BBS98} for $D=11$,
\cite{BNS03} for $D=10$ type IIA and \cite{D'ALST} for $D=10$ type IIB) where the
commutator of two supersymmetry transformations leaving invariant the supergravity
action would involve the PST (Pasti--Sorokin--Tonin) gauge transformations.}. A
practical way to pursue the above proposed procedure method to find the action up to
the terms of higher order in $\hat{\theta}(\xi)$ is to use a counterpart of the {\it
gauge completion} method (see \cite{CF80}), {\it but applied to the action itself}.
Namely, one makes an `infinitesimal' supersymmetry transformation in the action written
up to ${\cal O}(\hat{\theta}^{\wedge k})$ and recovers the next order in
$\hat{\theta}(\xi)$, ${\cal O}(\hat{\theta}^{\wedge (k+1)})$, by identifying
$\hat{\varepsilon}=\hat{\theta}(\xi)$; then one tries making such an action
supersymmetric up to order ${\cal O}(\hat{\theta}^{\wedge k})$ by modifying the
supersymmetry transformation rules of the $\hat{x}^\mu(\xi)$ and
$\hat{\theta}^\alpha(\xi)$ \footnote{To lowest  order $\delta_{\varepsilon}
\hat{x}^\mu(\xi)= - i \hat{\theta}\Gamma^a {\varepsilon}(\hat{x}) \,
e_a{}^{\mu}(\hat{x}) + {\cal O}(\hat{\theta})$, $\;
\delta_{\varepsilon}\hat{\theta}^\alpha(\xi)= -{\varepsilon}^\alpha(\hat{x}(\xi))+
{\cal O}(\hat{\theta})$.}. Such a procedure would also answer the question of whether a
fully supersymmetric (not gauge--fixed) interacting action $S_{SG}(e^a(x),
\psi^\alpha(x) , A_3(x)) + S_{M2}({e}^a(\hat{x}), A_3(\hat{x}); \hat{\theta}(\xi),
{\psi}^\alpha(\hat{x}))$, with $S_{M2}(\hat{e}^a, A_3(\hat{x});0,
{\psi}^\alpha(\hat{x}))= S^0_{M2}(\hat{e}^a, A_3(\hat{x}))$, exists formulated only in
terms of the physical fields of the supergravity multiplet and the superbrane
Goldstonions $\hat{x}(\xi)$ and $\hat{\theta}(\xi)$. As we discussed in this paper (and
may gathered from the results of \cite{deW+PP98}), the answer to this question is
affirmative up to second order in $\hat{\theta}(\xi)$. Notice that, if an obstruction
were found at some higher order in $\hat{\theta}$, it would pose an interesting
dilemma: whether such an obstruction is the result of a non-Lagrangian nature of the
equations of motion for the physical fields of the supergravity multiplet in the
interacting system, or whether it is the application of the above procedure to the
equations of motion for the physical fields of the supergravity multiplet that fails.
The second alternative would imply the impossibility of finding a fully supersymmetric
system of equations for the physical fields of the supergravity multiplet and the
superbrane Goldstone fields. Although at first glance this would look discouraging, it
might also point towards some hidden ingredients of M-theory.

\bigskip

\section*{Acknowledgments}

The authors thank Dima Sorokin for several valuable discussions.
We also wish to thank Eric Berghoeff, Sergio Ferrara, Toine Van
Proeyen, G. Moore and W. Siegel for  useful conversations at
different stages of this work. This paper has been partially
supported by the research grants BFM2002-03681 from the Ministerio
de Educaci\'on y Ciencia and EU FEDER funds, N 383 of the
Ukrainian State Fund for Fundamental Research, from Generalitat
Valenciana and by the EU network MRTN--CT--2004--005104
`Constituents, Fundamental Forces and Symmetries of the Universe'.

\end{document}